\documentclass[prb,superscriptaddress,amsfonts,amsmath,floatfix]{revtex4}
\usepackage{graphicx}
\usepackage{bm}
\usepackage{amsmath}
\usepackage{hyperref}

\begin{document}
\title{Approximate solution of wave propagation in transverse magnetic mode through a graded interface positive-negative using asymptotic iteration method\footnote{Presented at the ICTAP Bali, Indonesia, Oct 2014}}

\author{Andri S. Husein}
\email{decepticon1022@gmail.com}
\affiliation{Department of Physics, University of Sebelas Maret\\
Jalan Ir. Sutami 36 A, Surakarta 57126, Indonesia}
\author{C. Cari}
\affiliation{Department of Physics, University of Sebelas Maret\\
Jalan Ir. Sutami 36 A, Surakarta 57126, Indonesia}
\author{A. Suparmi}
\affiliation{Department of Physics, University of Sebelas Maret\\
Jalan Ir. Sutami 36 A, Surakarta 57126, Indonesia}
\author{Miftachul Hadi}
\affiliation{Department of Mathematics, Universiti Brunei Darussalam\\
Jalan Tungku Link BE1410, Gadong, Negara Brunei Darussalam}
\affiliation{Physics Research Centre, Indonesian Insitute of Sciences (LIPI)\\
Kompleks Puspiptek, Serpong, Tangerang 15314, Indonesia}
\affiliation{Department of Physics, School of Natural Sciences\\
Ulsan National Institute of Science and Technology (UNIST)\\
50, UNIST-gil, Eonyang-eup, Ulju-gun, Ulsan, South Korea}
\affiliation{Institute of Modern Physics, Chinese Academy of Sciences\\
509 Nanchang Rd., Lanzhou  730000, China}

\date{\today}
\begin{abstract}
We investigate the propagation of electromagnetic waves in transverse magnetic (TM) mode through the structure of materials interface that have permittivity or permeability profile graded positive-negative using asymptotic iteration method (AIM). As the optical character of materials, the permittivity and the permeability profiles have been designed from constant or hyperbolic functions. In this work we show the approximate solution of magnetic field distribution and the eight models of wave vector of materials  or interface positive-negative gradation.
\end{abstract}

\maketitle

\section{INTRODUCTION}
A new class of artificial composite material is negative refractive index metamaterials (NRM), which is also called left-handed metamaterials (LHM), has attracted the interest of many scientists for more than a decade. Veselago \cite{1} who first published a review
of theoretical NRM for more than four decades ago. Then Pendry translate these concepts into practical applications \cite{3,4,5} which then attracted great attention on this topic. Experimental confirmation is given by Smith et al \cite{6} and Shelby et al \cite{17}. 

The concept of NRM is quite simple. The function of the material is more likely to be determined by the structure rather than by chemical composition. This is a new class of materials that goes beyond conventional materials because it has the effect of which has never been observed before \cite{14}.

NRM has structure in the order of subwavelengths and is able to provide negative refractive index in a certain wavelength range. The artificial structure contains negative permeability obtained from the double split-ring resonator and negative permittivity obtained from nano-wires \cite{6}. The initial idea of Veselago derived from the dispersion equation which expresses the interaction of electromagnetic waves with a dielectric. Dielectric response to the presence of an electromagnetic wave is expressed by the basic characteristics of the two
quantities, i.e. permittivity, $\varepsilon$, and permeability, $\mu$. That is because only these two quantities that appear in the dispersion equation \cite{1}. 

Using a diagram of $\varepsilon -\mu$, the material properties can be grouped into four classes i.e. \cite{1}:
(i) Ordinary materials ($\varepsilon>0,~\mu>0$); (ii) Materials with negative permittivity and positive permeability ($\varepsilon<0,~\mu>0$); (iii) NRM ($\varepsilon<0,~\mu<0$) and (iv) Materials with positive permittivity and negative permeability ($\varepsilon>0,~\mu<0$).

Analysis of the wave propagation through the NRM structures most commonly done with modeling and numerical simulation, in particular using the finite difference time domain method (FDTD). FDTD method has been widely accepted as one of the popular numerical methods in computational electromagnetic \cite{2}. At present, the method used to design and to obtain insight into the physical characteristics of
electromagnetic NRM \cite{13}. NRM structures with graded refractive index interface negative-positive has been solved analytically by M. Dalarsson for the special case in which its index profile is a linear or exponential function \cite{15}. 

Actually, the real structure of any material contains positive and negative refractive index and at the same time tend to have a gradation profile \cite{10}. NRM gradation index has been widely studied. Ramakrishna describes an NRM lenses composed of media with a gradient index \cite{9}. Calculation of the transmittance at an exponential gradation structure NRM has been given by N. Dalarsson \cite{10}. Analytical solutions obtained from hypergeometry differential equations compared with the numerical solution using the transfer matrix method (TMM) seem to have a good pattern match. 

Electromagnetic wave equation for inhomogeneous medium can be converted into a second order homogeneous ordinary differential equation which is linear in one dimensional case. Thus, the analytical solution of Maxwell's equations can be solved in this domain. Homogeneous second order ordinary differential equations appeared widely in the literature and there are several ways to obtain an exact solution such as using methods of hypergeometry, supersymmetry quantum mechanics \cite{7}, Nikiforov-Uvarov (NU) \cite{18}, Romanovski polynomials \cite{19,20,21} and which has recently been used is asymptotic iteration method (AIM) to solve the eigenvalue problem \cite{8} and electromagnetic wave propagation in inhomogeneous ordinary material \cite{12}. 

Exact solution of field distribution (eigenfunction) and energy (eigenvalue) can be obtained using the AIM if the Hamiltonian of the system is shaped like a harmonic oscillator \cite{8}. In this paper we try to apply the AIM in the case of inhomogeneous materials where permittivity or permeability has a positive-negative gradation profile. Here we restrict the calculation just on the approximate solution.

\section{METHOD}
In this section we present mathematical background to manage calculations of light propagation in the interface of NRM with positive-negative
gradation. We will discuss this study in three subsections: (i) we manipulate Maxwell's equations to construct the wave equation in inhomogeneous materials; (ii) we apply AIM approach to obtain solutions of second order homogeneous differential equation; (iii) we apply AIM to obtain approximate solution to the eight models of wave vector of materials.

\subsection{Field Equation}
To simplify the calculations, we set the $xz$-plane as the plane of ray incident, so $\partial/\partial y = 0$. In addition, the optical characteristics of $\varepsilon(x)$ and $\mu(x)$ are considered as functions of material thickness, $x$. Using Maxwell's equations and considering the geometry of the material, we obtain a second order differential equation of electromagnetic waves in inhomogeneous materials on mode Transverse Magnetic (TM). Actually, in two-dimensional study, there are two independent modes i.e. TM mode, the group of fields:
$\left\{E_x,~E_y,~H_z\right\},~\left\{E_y,~E_z,~H_x\right\},~\left\{E_z,~E_x,~H_y\right\}$ and Transverse Electric (TE) mode, the group of fields: $\left\{H_x,~H_y,~E_z\right\},~\left\{H_y,~H_z,~E_x\right\},~\left\{H_z,~H_x,~E_y\right\}$. 

In principle, these two modes (TM and TE) are similar to each other, so we only need to do one calculation, e.g. the TM mode. Now, let's start by defining the form of plane waves which is a special solution of the wave equation
\begin{eqnarray}\label{1}
H = H(x)~e^{i(\omega t-kz)}
\end{eqnarray}
and from Maxwell's equations is known that 
\begin{eqnarray}\label{2}
i\omega\mu(x)~H = -\nabla\times E;~~~ i\omega\varepsilon(x)~E = \nabla\times H 
\end{eqnarray}
Consider the group of fields in the TM mode which we use i.e. $\left\{E_z,~E_x,~H_y\right\}$. From eq.(\ref{1}) and eq.(\ref{2}) we obtain
\begin{eqnarray}\label{3}
\frac{\partial^2H(x)}{\partial x^2} -\frac{\varepsilon'(x)}{\varepsilon(x)}\frac{\partial H(x)}{\partial x} +[\omega^2\mu(x)\varepsilon(x)-k^2]H(x) = 0
\end{eqnarray}
where in eq.(\ref{3}) above, the prime symbol $(')$ indicates the operation of differentiation with respect to $x$. This second order differential equation does not always have the exact solution to a number of functions $\varepsilon(x)$ and $\mu(x)$ . In the next subsection, we will introduce the approach that is needed to achieve that goal.

\subsection{Asymptotic Iteration Method}
Asymptotic iteration method aims to obtain the exact solution of a second order differential equation
which has the following form
\begin{eqnarray}\label{4}
Y''(x) -\lambda_0(x)Y'(x) -s_0(x)Y(x) = 0
\end{eqnarray}
where $\lambda_0(x)\neq0$ and $s_0(x)$ is the differential equation coefficients and they are so well-defined functions which are differentiable. To find the general solution, eq.(\ref{4}) needs to be differentiable in $(m + 1)$ times and $(m + 2)$ times, where $m=1,2,3,..$ is the number of iterations. Then one can obtain
\begin{eqnarray}\label{5}
Y^{(m+1)}(x) -\lambda_{m-1}(x)Y'(x) -s_{m-1}(x)Y(x) = 0
\end{eqnarray}
\begin{eqnarray}\label{6}
Y^{(m+2)}(x) -\lambda_{m}(x)Y'(x) -s_{m}(x)Y(x) = 0
\end{eqnarray}
where
\begin{eqnarray}\label{7}
\lambda_m(x)
&=& \lambda'_{m-1}(x) +\lambda_{m-1}(x)\lambda_0(x) +s_{m-1}(x)
\end{eqnarray}
\begin{eqnarray}\label{8}
s_m(x) =s'_{m-1}(x) +s_0(x)\lambda_{m-1}(x)
\end{eqnarray}
which are also called recursive relations of eq.(\ref{4}). The calculation of the ratio of derivatives of $(m + 2)$ and $(m + 1)$ gives
\begin{eqnarray}\label{9}
\frac{d}{dx}\ln\left[Y^{(m+1)}(x)\right]
&=& \frac{Y^{(m+2)}(x)}{Y^{(m+1)}(x)} \nonumber\\
&=& \frac{\lambda_m[Y'(x) +(s_m/\lambda_m)Y(x)]}{\lambda_{m-1}[Y'(x) +(s_{m-1}/\lambda_{m-1})Y(x)]}
\end{eqnarray}
Furthermore, with the introduction of the asymptotic aspects of this iteration method, i.e. by assuming that the value of $m$ is large enough, we obtain
\begin{eqnarray}\label{10}
\frac{s_m(x)}{\lambda_m(x)} -\frac{s_{m-1}(x)}{\lambda_{m-1}(x)} \equiv \alpha(x)
\end{eqnarray}
So eq.(\ref{9}) can be written as
\begin{eqnarray}\label{11}
\frac{d}{dx}\ln[Y^{(m+1)}(x)] = \frac{\lambda_m(x)}{\lambda_{m-1}(x)}
\end{eqnarray}
Substitute eq.(\ref{7}) into eq.(\ref{11}) and using eq.(\ref{10}), we obtain
\begin{eqnarray}\label{12}
Y^{(m+1)}(x) =C_0\lambda_{m-1}(x)~\exp\left\{\int[\alpha(x) +\lambda_0(x)]~dx \right\}
\end{eqnarray}
where $C_0$ is the constant of integration. Substitute eq.(\ref{12}) into eq.(\ref{5}) and solve for $Y(x)$, then we obtain the general solution eq.(\ref{4}) as
\begin{eqnarray}\label{13}
Y(x) = \exp\left[-\int\alpha(x)dx\right]\times\left(C_1 +C_0\int\exp\left\{\int[2\alpha(x) +\lambda_0(x)]dx\right\}dx\right)
\end{eqnarray}
where $C_1$ is the constant of integration. 

Eigenvalue of energy is obtained from the combination of termination conditions, eq.(\ref{10}) and eqs.(\ref{7})-(\ref{8}).
\begin{eqnarray}\label{14}
\lambda_m(x)s_{m-1}(x) -\lambda_{m-1}(x)s_m(x)=0;~~~m=1,2,3
\end{eqnarray}
In general, if the problem expressed by eq.(\ref{4}) has $\lambda_0(x)$ and $s_0(x)$ which are forming a constant function, then the general solution of eq.(\ref{4}) can directly follow to eq.(\ref{13}). The most frequent case is that $\lambda_0(x)$ or $s_0(x)$ is not a constant function. In such cases, the exact solution eq.(\ref{4}) requires a special technique that can be found in \cite{8} \cite{11,12} \cite{16}. 

Here, we will apply the AIM to the case where the electromagnetic optical character changed gradually. Our main attention is the solution domain where the positive-negative interface occurs and how to develop appropriate approaches. As an overview of approaches that we use here, let us look at the following example. \\\\
Example:\\
Optical characteristics of vacuum space is defined
as follows
\begin{eqnarray}\label{15}
\varepsilon(x)=\varepsilon_0;~~~\mu(x)=\mu_0
\end{eqnarray}
with $\varepsilon_0(x)$ and $\mu_0$ are respectively the permittivity and permeability constants of vacuum space, and $\mu_0\varepsilon_0=1/c^2$. Substitute eq.(\ref{15}) into eq.(\ref{3}), we obtain
\begin{eqnarray}\label{16}
\frac{d^2H(x)}{dx^2} +(\omega^2/c^2 -k^2)H(x)=0
\end{eqnarray}
Furthermore, if we let
\begin{eqnarray}\label{17}
H(x) = e^{-x^2/2}\psi(x)
\end{eqnarray}
where $\psi(x)$ needs to be determined using an iterative procedure. Substitute eq.(\ref{17}) into eq.(\ref{16}) gives
\begin{eqnarray}\label{18}
\frac{d^2\psi(x)}{dx^2} -\lambda_0(x)\frac{d\psi(x)}{dx} +s_0(x)\psi(x)=0
\end{eqnarray}
with
\begin{eqnarray}\label{19}
\lambda_0(x)=2x;~~~s_0(x)=(\omega^2/c^2 -k^2)-1+x^2
\end{eqnarray}
In the solution domain close to zero, eq.(\ref{19}) can be approximated by MacLaurin series. The approach is carried by taking the first two terms of the MacLaurin series, giving
\begin{eqnarray*}
\lambda_0(x)
&=& \lambda_0(0) +\lambda_0'(0)x +\frac{\lambda_0''(0)}{2!}x^2 +\frac{\lambda_0'''(0)}{3!}x^3 +... \nonumber\\
&=& 0 +2x +0 + ... \nonumber\\
&=& 2x
\end{eqnarray*}
\begin{eqnarray}\label{20}
s_0(x)
&=& s_0(0) +s_0'(0)x +\frac{s_0''(0)}{2!}x^2 +\frac{s_0'''(0)}{3!}x^3 +... \nonumber\\
&=& (\omega^2/c^2 -k^2) -1 +0\times x +\frac{2}{2!}x^2 +0 +... \nonumber\\
&=& (\omega^2/c^2 -k^2) -1
\end{eqnarray}
Substitute eq.(\ref{20}) into eq.(\ref{18}), we obtain
\begin{eqnarray}\label{21}
\frac{d^2\psi(x)}{dx^2} -2x\frac{d\psi(x)}{dx} +[(\omega^2/c^2 -k^2)-1]\psi(x)=0
\end{eqnarray}
at $-\delta\leq x\leq \delta$, with $\delta=1$. Furthermore, using eq.(\ref{14}), it can be obtained
\begin{eqnarray}\label{22}
\lambda_ms_{m-1}-\lambda_{m-1}s_m
&=&\Pi_{n=0}^{m}[(\omega^2/c^2 -k^2 -1)-2n];~~~m=1,2,3,4,...
\end{eqnarray}
According to these conditions, the eigenvalues of energy can be obtained as follows
\begin{eqnarray}\label{23}
\frac{\omega^2}{c^2} -k^2 -1 =2n\rightarrow k_n=\sqrt{\frac{\omega^2}{c^2} -(2n+1)};~~~n=0,1,2,3,...
\end{eqnarray}
Using eq.(\ref{23}), eq.(\ref{21}) can be written as
\begin{eqnarray}\label{24}
\frac{d^2\psi(x)}{dx^2} -2x\frac{d\psi(x)}{dx}~2n\psi(x) = 0;~~~n=0,1,2,3,...
\end{eqnarray}
Eq.(\ref{24}) known as Hermite differential equation which has a solution in a polynomial of order $n$. The approximate solution of eq.(\ref{16}) is
\begin{eqnarray}\label{25}
H_n(x) = e^{-x^2/2}(-1)^ne^{x^2}~\frac{d}{dx^n}(e^{-x^2});~~~n=0,1,2,3,..
\end{eqnarray}
at $-\delta\leq x\leq \delta$, with $\delta = 1$.
Using similar techniques, we develop a method to solve the eight models of interface positive-negative gradation as follows.

\subsubsection{1st Model}
Define the character of the optical interface gradation as follows
\begin{eqnarray*}
\varepsilon(x)=\varepsilon_0[csch(\rho x) -\coth(\rho x)]
\end{eqnarray*}
\begin{eqnarray}\label{26}
\mu(x) = -\mu_0[csch(\rho x) +\coth(\rho x)]
\end{eqnarray}
with $\rho$ is the gradation parameters. If we let $u=\rho x$, substitute eq.(\ref{26}) into eq.(\ref{3}) gives
\begin{eqnarray}\label{27}
\frac{d^2H}{du^2} -\frac{csch(u)}{\rho}\frac{dH}{du} +\frac{1}{\rho^2}\left(\frac{\omega^2}{c^2} -k^2\right)H = 0
\end{eqnarray}
In the solution domain close to zero, using the MacLaurin series for the first three terms, then we obtain
\begin{eqnarray}\label{28}
csch(u)=\frac{1}{u}
\end{eqnarray}
However, eq.(\ref{28}) is not defined at $u = 0$, so we need an assumption. If $\sigma$ is the value of $u$ that close to zero then using eq.(\ref{28}) it can be obtained $csch(\sigma) = 1/\sigma$, and $csch(-\sigma) = -1/\sigma$. Furthermore, at intervals of $-\sigma\leq u\leq \sigma$, eq.(\ref{28}) is considered to have a linear function form which connects the points $(-\sigma,-1/\sigma)$ and $(\sigma,1/\sigma)$, i.e.
\begin{eqnarray}\label{29}
f(u) = \frac{1}{\sigma^2}u
\end{eqnarray}
Using (\ref{29}) and letting $z=u/\sqrt{2\rho\sigma^2}$, eq.(\ref{27}) can be written as
\begin{eqnarray}\label{30}
\frac{d^2H}{dz^2} -2z\frac{dH}{dz} +\left(\frac{2\sigma^2\omega^2}{\rho c^2} -\frac{2\sigma^2}{\rho}k^2\right)H = 0
\end{eqnarray}
at $-\sigma\leq u\leq\sigma$, with $\sigma=1$. Furthermore, using the eq.(\ref{14}), it can be obtained
\begin{eqnarray}\label{31}
\lambda_ms_{m-1} -\lambda_{m-1}s_m
&=&\Pi_{n=0}^{m}\left[\left(\frac{2\sigma^2\omega^2}{\rho c^2} -\frac{2\sigma^2}{\rho}k^2\right) -2n\right]
\end{eqnarray}
According to these conditions, the eigenvalues of energy can be obtained as follows
\begin{eqnarray}\label{32}
\left(\frac{2\sigma^2\omega^2}{\rho c^2} -\frac{2\sigma^2}{\rho}k^2\right) = 2n\rightarrow k_n=\sqrt{\frac{\omega^2}{c^2} -\frac{\rho n}{\sigma^2}}
\end{eqnarray}
Using eq.(\ref{32}), eq.(\ref{30}) can be rewritten as
\begin{eqnarray}\label{33}
\frac{d^2H}{dz^2} -2z\frac{dH}{dz} +2nH = 0;~~~n=0,1,2,3,...
\end{eqnarray}
Eq.(\ref{33}) is known as Hermite differential equation. The approximate solution of eq.(\ref{27}) is
\begin{eqnarray}\label{34}
H_n(z) = (-1)^ne^{z^2}\frac{d^n}{dz^n}(e^{-z^2});~~~n=0,1,2,3,...
\end{eqnarray}

\subsubsection{2nd Model}
Define the character of the optical interface gradation as follows
\begin{eqnarray*}
\varepsilon(x) = -\varepsilon_0[csch(\rho x) +\coth(\rho x)]
\end{eqnarray*}
\begin{eqnarray}\label{35}
\mu(x) = \mu_0[csch(\rho x) -\coth(\rho x)]
\end{eqnarray}
If we let $u=\rho x$, substitute eq.(\ref{35}) into eq.(\ref{3}) gives
\begin{eqnarray}\label{36}
\frac{d^2H}{du^2} +\frac{csch(u)}{\rho}\frac{dH}{du} +\frac{1}{\rho^2}\left(\frac{\omega^2}{c^2} -k^2\right)H = 0
\end{eqnarray}
Compare to eq.(\ref{30}), eq.(\ref{36}) differs only the sign in the middle term of eq.(\ref{30}). So, in the same way we can obtain
\begin{eqnarray}\label{37}
\frac{d^2H}{dz^2} +2z\frac{dH}{dz} +\left(\frac{2\sigma^2\omega^2}{\rho c^2} -\frac{2\sigma^2}{\rho}k^2\right)H = 0
\end{eqnarray}
at $-\sigma\leq u\leq\sigma$, with $\sigma=1$. Furthermore, using eq.(\ref{14}), we obtain
\begin{eqnarray}\label{38}
\lambda_ms_{m-1}-\lambda_{m-1}s_m =\Pi_{n=0}^{m}\left[\left(\frac{2\sigma^2\omega^2}{\rho c^2} -\frac{2\sigma^2}{\rho}k^2\right) +2n\right]
\end{eqnarray}
According to these conditions, the eigenvalues of energy can be obtained as follows
\begin{eqnarray}\label{39}
\left(\frac{2\sigma^2\omega^2}{\rho c^2} -\frac{2\sigma^2}{\rho}k^2\right) = -2n\rightarrow k_n=\sqrt{\frac{\omega^2}{c^2} +\frac{\rho n}{\sigma^2}}
\end{eqnarray}
Using eq.(\ref{39}), eq.(\ref{37}) can be written as
\begin{eqnarray}\label{40}
\frac{d^2H}{dz^2} +2z\frac{dH}{dz} -2nH=0;~~~n=0,1,2,3,..
\end{eqnarray}
Eq.(\ref{40}) is known as Differential Equations of Complementary Error Function. The approximate solution of eq.(\ref{36}) is
\begin{eqnarray}\label{41}
H_n(z)=A~erf~c_n(z) +B~erf~c_n(-z);~~~n=0,1,2,3,..
\end{eqnarray}
with $A$ and $B$ are constant.
\begin{eqnarray}\label{42}
erf~c_n(z)
&=& \int_z^\infty ...\int_z^\infty erf(z)~dz\nonumber\\
&=& 2^{-n}e^{-z^2}\left[\frac{{}_1F_1[1/2(n+1);1/2;z^2]}{\Gamma[1+(1/2)n]} -\frac{2z~{}_1F_1[1 +(1/2)n;3/2;z^2]}{\Gamma[1/2(n+1)]}\right]
\end{eqnarray}
is an integral $n$ times the Complementary Error Function, and
\begin{eqnarray}\label{43}
{}_1F_1(a;b;z)
&=& 1 +\frac{a}{b}z +\frac{a(a+1)}{b(b+1)}\frac{z^2}{2!} +\frac{a(a+1)(a+2)}{b(b+1)(b+2)}\frac{z^3}{3!} +...\nonumber\\
&=& \Sigma_{k=0}^{\infty}\frac{(a)_kz^k}{(b)_kk!}
\end{eqnarray}
is the confluent hypergeometric function of first kind, whereas $\Gamma(z)$ is the gamma function.

\subsubsection{3rd Model}
Define the character of the optical interface gradation as follows 
\begin{eqnarray}\label{44}
\varepsilon(x)
&=& \varepsilon_0;~~~\mu(x)=\mu_0[csch(\rho x) -\coth(\rho x)]
\end{eqnarray}
If we let $z=\frac{1}{2}\rho x$, substitute eq.(\ref{44}) into eq.(\ref{3}) we obtain
\begin{eqnarray}\label{45}
\frac{d^2H}{dz^2} -\frac{4}{\rho^2}\left(\frac{\omega^2}{c^2}\tanh(z)+k^2\right)H = 0
\end{eqnarray}
In the solution domain close to zero, it can be approached, $\tanh(z)=z$, in order to obtain
\begin{eqnarray}\label{46}
\frac{d^2H}{dz^2} -\left(\frac{4\omega^2}{\rho^2c^2} +\frac{4}{\rho^2}k^2\right)H = 0
\end{eqnarray}
at $-\delta\leq z\leq\delta$ with $\delta = 1$. If we let $s=\left(4\omega^2/\rho^2c^2\right)^{1/3}z$, and
\begin{eqnarray}\label{47}
H(s) = e^{\frac{s^3}{6}-\frac{s^2}{2}}\psi(s)
\end{eqnarray}
with $\psi(s)\psi(x)$ needs to be determined using an iterative procedure, we obtain
\begin{eqnarray}\label{48}
\frac{d^2\psi}{ds^2} -\lambda_0(s)\frac{d\psi}{ds} -s_0(s)\psi =0
\end{eqnarray}
with
\begin{eqnarray}\label{49}
\lambda_0(s) = 2s - s^2
\end{eqnarray}
and
\begin{eqnarray}\label{50}
s_0(s) = 1 +\frac{4}{\rho^2}\left(\frac{\rho^2c^2}{4\omega^2}\right)^{2/3}k^2 -s^2 +s^3 -\frac{1}{4}s^4
\end{eqnarray}
Using the MacLaurin series for the first two terms, eq.(\ref{49}) and eq.(\ref{50}) can be respectively reduced to
\begin{eqnarray}\label{51}
\lambda_0(s) = 2s
\end{eqnarray}
and
\begin{eqnarray}\label{52}
s_0(s) = 1 +\frac{4}{\rho^2}\left(\frac{\rho^2c^2}{4\omega^2}\right)^{2/3}k^2
\end{eqnarray}
Substitute eq.(\ref{51}) and eq.(\ref{52}) into eq.(\ref{48}), we obtain
\begin{eqnarray}\label{53}
\frac{d^2\psi}{ds^2} -2s\frac{d\psi}{ds} -\left[1 +\frac{4}{\rho^2}\left(\frac{\rho^2c^2}{4\omega^2}\right)^{2/3}k^2\right]\psi = 0
\end{eqnarray}
Furthermore, using eq.(\ref{14}), it can be obtained
\begin{eqnarray}\label{54}
\lambda_ms_{m-1} -\lambda_{m-1}s_m = \Pi_{n=0}^{m}\left\{\left[1 +\frac{4}{\rho^2}\left(\frac{\rho^2c^2}{4\omega^2}\right)^{2/3}k^2\right]+2n\right\};~~~m=1,2,3,4,..
\end{eqnarray}
According to these conditions, the eigenvalues of energy can be obtained as follows
\begin{eqnarray}\label{55}
1 +\frac{4}{\rho^2}\left(\frac{\rho^2c^2}{4\omega^2}\right)^{2/3}k^2 = -2n\rightarrow k_n=\frac{i\rho}{2}\left(\frac{2\omega}{\rho c}\right)^{2/3}\sqrt{2n +1}
\end{eqnarray}
Using eq.(\ref{55}), eq.(\ref{53}) can be rewritten as
\begin{eqnarray}\label{56}
\frac{d^2\psi}{ds^2} -2s\frac{d\psi}{ds} +2n\psi = 0;~~~n=0,1,2,3,...
\end{eqnarray}
Eq.(\ref{56}) is known as Hermite differential equation. The approximate solution of eq.(\ref{45}) is
\begin{eqnarray}\label{57}
H_n(s) = e^{\frac{s^3}{6}-\frac{s^2}{2}}(-1)^ne^{s^2}\frac{d^n}{ds^n}\left(e^{-s^2}\right);~~~n=0,1,2,3,...
\end{eqnarray}

\subsubsection{4th Model}
Define the character of the optical interface gradation as follows
\begin{eqnarray}\label{58}
\varepsilon(x) 
&=& \varepsilon_0[csch(\rho x)-\coth(\rho x)];~~~\mu(x)=\mu_0
\end{eqnarray}
If we let $z=\frac{1}{2}\rho x$ and substitute eq.(\ref{58}) into eq.(\ref{3}) we obtain
\begin{eqnarray}\label{59}
\frac{d^2H}{dz^2} -\frac{csch(z)~sech(z)}{\rho}\frac{dH}{dz} -\frac{4}{\rho^2}\left(\frac{\omega^2}{c^2}\tanh(z) +k^2\right)H = 0
\end{eqnarray}
In the solution domain close to zero, using the MacLaurin series for the first three terms, we obtain
\begin{eqnarray}\label{60}
\tanh(z)=z
\end{eqnarray}
\begin{eqnarray}\label{61}
csch(z)~sech(z)=\frac{1}{z}
\end{eqnarray}
However, eq.(\ref{61}) is not defined at $z = 0$, so it needs an assumption. If $\sigma$ is the value of $z$ that close to zero then using eq.(\ref{61}) it can be obtained $csch(\sigma)sech(\sigma) = 1/\sigma$ and $csch(-\sigma)sech(-\sigma) = -1/\sigma$. Furthermore, at intervals of $-\sigma\leq z\leq\sigma$, eq.(\ref{61}) is considered to have a linear function form which is connecting the points $(-\sigma, -1/\sigma)$ and $(\sigma,1/\sigma)$, i.e.
\begin{eqnarray}\label{62}
f(u) = \frac{1}{\sigma^2}z
\end{eqnarray}
Using eq.(\ref{60}) and eq.(\ref{62}), eq.(\ref{59}) can be rewritten as
\begin{eqnarray}\label{63}
\frac{d^2H}{dz^2} -\frac{z}{\sigma^2\rho}\frac{dH}{dz} -\frac{4}{\rho^2}\left(\frac{\omega^2}{c^2}z +k^2\right)H = 0
\end{eqnarray}
at $-\sigma\leq z\leq\sigma$ with $\sigma=1$. If we let $s=(4\omega^2/\rho^2c^2)^{1/3}z$, and
\begin{eqnarray}\label{64}
H(s) = e^{s^3/6}\psi(s)
\end{eqnarray}
with $\psi(s)$ needs to be determined using an iterative procedure, then substitute eq.(\ref{64}) into eq.(\ref{63}) we obtain
\begin{eqnarray}\label{65}
\frac{d^2\psi}{ds^2} -\lambda_0(s)\frac{d\psi}{ds} -s_0(s)\psi = 0
\end{eqnarray}
with
\begin{eqnarray}\label{66}
\lambda_0(s) =\frac{1}{\sigma^2\rho}\left(\frac{\rho^2c^2}{4\omega^2}\right)^{2/3}s-s^2
\end{eqnarray}
and
\begin{eqnarray}\label{67}
s_0(s)=\frac{4}{\rho^2}\left(\frac{\rho^2 c^2}{4\omega^2}\right)^{2/3}k^2 +\frac{1}{2\sigma^2\rho}\left(\frac{\rho^2c^2}{4\omega^2}\right)^{2/3}s^3 -\frac{1}{4}s^4.
\end{eqnarray}
Using the MacLaurin series for the first two terms, eq.(\ref{66}) and eq.(\ref{67}) respectively can be reduced to
\begin{eqnarray}\label{68}
\lambda_0(s) = \frac{1}{\sigma^2\rho}\left(\frac{\rho^2c^2}{4\omega^2}\right)^{2/3}s
\end{eqnarray}
and
\begin{eqnarray}\label{69}
s_0(s) = \frac{4}{\rho^2}\left(\frac{\rho^2 c^2}{4\omega^2}\right)^{2/3}k^2
\end{eqnarray}
Substitute eq.(\ref{68}) and eq.(\ref{69}) into eq.(\ref{65}), we obtain
\begin{eqnarray}\label{70}
\frac{d^2\psi}{ds^2} -\frac{1}{\sigma^2\rho}\left(\frac{\rho^2 c^2}{4\omega^2}\right)^{2/3}s\frac{d\psi}{ds} -\frac{4}{\rho^2}\left(\frac{\rho^2 c^2}{4\omega^2}\right)^{2/3}k^2\psi = 0
\end{eqnarray}
Furthermore, we let $r=\left(\rho^2c^2/4\omega^2\right)^{1/3}s/\sqrt{2\sigma^2\rho}$, eq.(\ref{70}) can be written as
\begin{eqnarray}\label{71}
\frac{d^2\psi}{dr^2} -2r\frac{d\psi}{dr} -\frac{8\sigma^2}{\rho}k^2\psi = 0
\end{eqnarray}
Then, using eq.(\ref{14}), we obtain
\begin{eqnarray}\label{72}
\lambda_ms_{m-1}-\lambda_{m-1}s_m =\Pi_{n=0}^{m}\left(\frac{8\sigma^2}{\rho}k^2 +2n\right);~~~m=1,2,3,4,..
\end{eqnarray}
According to these conditions, the eigenvalues of energy can be obtained as follows
\begin{eqnarray}\label{73}
\frac{8\sigma^2}{\rho}k^2 = -2n\rightarrow k_n=\frac{i}{2\sigma}\sqrt{\rho n};~~~n=0,1,2,3,..
\end{eqnarray}
Using eq.(\ref{73}), then eq.(\ref{71}) can be written as
\begin{eqnarray}\label{74}
\frac{d^2\psi}{dr^2} -2r\frac{d\psi}{dr} +2n\psi = 0;~~~n=0,1,2,3,...
\end{eqnarray}
Eq.(\ref{74}) known as Hermite differential equation. The approximate solution of eq.(\ref{59}) is
\begin{eqnarray}\label{75}
H_n(r) = e^{\frac{4\sqrt{2\sigma\omega^2}}{3\sqrt{\rho c^2}}r^3}(-1)^ne^{r^2}\frac{d^n}{dr^n}(e^{-r^2});~~~n=0,1,2,3,...
\end{eqnarray}

\subsubsection{5th Model}
Define the character of the optical interface gradation as follows
\begin{eqnarray}\label{76}
\varepsilon(x) = \varepsilon_0;~~~\mu(x)=-\mu_0[csch(\rho x) +\coth(\rho x)]
\end{eqnarray}
If we let $u=\rho x$, and substitute eq.(\ref{76}) into eq.(\ref{3}), we obtain
\begin{eqnarray}\label{77}
\frac{d^2H}{du^2} -\frac{1}{\rho^2}\left\{\frac{\omega^2}{c^2}[csch(u) +\coth(u)] +k^2\right\}H=0
\end{eqnarray}
In the solution domain close to zero, using the MacLaurin series for the first three terms, it gives
\begin{eqnarray}\label{78}
csch(u)~\coth(u) = \frac{2}{u} +\frac{u}{2}
\end{eqnarray}
However, eq.(\ref{78}) is not defined at $u = 0$, so it needs an assumption. If $\sigma$ is the value of $u$ that
close to zero then using eq.(\ref{78}) we obtain
$csch(\sigma) + \coth(\sigma) = (\sigma^2 + 4)/2\sigma$ and
$csch(-\sigma) + \coth(-\sigma) = -(\sigma^2 + 4)/2\sigma$. Furthermore, at intervals of $-\sigma\leq u\leq \sigma$, eq.(\ref{78}) is considered to have a linear function form which is connecting the points $[-\sigma, -(\sigma^2 + 4)/2\sigma]$ and $[\sigma,(\sigma^2+4)/2\sigma]$, i.e.
\begin{eqnarray}\label{79}
f(u) = \frac{\sigma^2+4}{2\sigma^2}u
\end{eqnarray}
Using eq.(\ref{79}), then eq.(\ref{77}) can be written as
\begin{eqnarray}\label{80}
\frac{d^2H}{du^2} -\frac{1}{\rho^2}\left(\frac{\omega^2}{c^2}\frac{\sigma^2+4}{2\sigma}u +k^2\right)H = 0
\end{eqnarray}
at $-\sigma\leq u\leq\sigma$ with $\sigma=1$. If we let $\beta=\omega^2(\sigma^2 +4)/2c^2\rho^2\sigma^2$, $z=\beta^{1/3}u$ and
\begin{eqnarray}\label{81}
H(z) = e^{\frac{z^3}{6}-\frac{z^2}{2}}\psi(z)
\end{eqnarray}
with $\psi(z)$ needs to be determined using an iterative procedure, then substitute eq.(\ref{81}) into eq.(\ref{80}) we obtain
\begin{eqnarray}\label{82}
\frac{d^2\psi}{dz^2} -\lambda_0(z)\frac{d\psi}{dz} -s_0(z)\psi = 0
\end{eqnarray}
with
\begin{eqnarray}\label{83}
\lambda_0(z) = 2z -z^2
\end{eqnarray}
and
\begin{eqnarray}\label{84}
s_0(z) = 1 +\frac{k^2}{\beta^{2/3}\rho^2} -z^2 +z^3 -\frac{1}{4}z^4
\end{eqnarray}
Using the MacLaurin series for the first two terms, eq.(\ref{83}) and eq.(\ref{84}) respectively can be reduced to
\begin{eqnarray}\label{85}
\lambda_0(z) = 2z
\end{eqnarray}
and
\begin{eqnarray}\label{86}
s_0(z) = 1 +\frac{k^2}{\beta^{2/3}\rho^2}
\end{eqnarray}
Substitute eq.(\ref{85}) and eq.(\ref{86}) into eq.(\ref{82}), we obtain
\begin{eqnarray}\label{87}
\frac{d^2\psi}{dz^2} -2z\frac{d\psi}{dz} -\left(1 +\frac{k^2}{\beta^{2/3}\rho^2}\right)\psi = 0
\end{eqnarray}
Then, using eq.(\ref{14}), we obtain
\begin{eqnarray}\label{88}
\lambda_ms_{m-1} -\lambda_{m-1}s_m =\Pi_{n=0}^{m}\left(1 +\frac{k^2}{\beta^{2/3}\rho^2} +2n\right);~~~m=1,2,3,4,..
\end{eqnarray}
According to these conditions, the  eigenvalues of energy can be obtained as follows
\begin{eqnarray}\label{89}
1 +\frac{k^2}{\beta^{2/3}\rho^2} = -2n\rightarrow k_n=i\beta^{2/3}\rho\sqrt{2n+1};~~~n=0,1,2,3,..
\end{eqnarray}
Using eq.(\ref{89}), then eq.(\ref{87}) can be written as
\begin{eqnarray}\label{90}
\frac{d^2\psi}{dz^2} -2z\frac{d\psi}{dz} +2n\psi = 0;~~~n=0,1,2,3,..
\end{eqnarray}
Eq.(90) known as Hermite differential equation. The approximate solution of eq.(\ref{77}) is
\begin{eqnarray}\label{91}
H_n(z) = e^{\frac{z^3}{6}-\frac{z^2}{2}}(-1)^ne^{z^2}\frac{d^n}{dz^n}\left(e^{-z^2}\right);~~~n=0,1,2,3,...
\end{eqnarray}

\subsubsection{6th Model}
Define the character of the optical interface gradation as follows 
\begin{eqnarray}\label{92}
\varepsilon(x) 
&=& -\varepsilon_0[csch(\rho x)+\coth(\rho x)];~~~\mu(x) = \mu_0
\end{eqnarray}
If we let $u =\rho x$ and substitute eq.(\ref{92}) into eq.(\ref{3}) gives
\begin{eqnarray}\label{93}
\frac{d^2H}{du^2} +\frac{csch(u)}{\rho}\frac{dH}{du} -\frac{1}{\rho^2}\left\{\frac{\omega^2}{c^2}[csch(u)+\coth(u)]+k^2\right\}H
\end{eqnarray}
In the solution domain close to zero, using the MacLaurin series for the first three terms, functions of $csch(u)$ and $csch(u) + \coth(u)$ can be approximated by the MacLaurin series as eq.(\ref{28}) and eq.(\ref{78}), so that eq.(\ref{93}) can be reduced to
\begin{eqnarray}\label{94}
\frac{d^2H}{du^2} +\frac{u}{\sigma^2\rho}\frac{dH}{du} -\frac{1}{\rho^2}\left[\frac{\omega^2}{c^2}\left(\frac{\sigma^2+4}{2\sigma^2}\right)u+k^2\right]H = 0
\end{eqnarray}
at $-\sigma\leq u\leq\sigma$ with $\sigma=1$. Furthermore, if we let $\beta = \omega^2(\sigma^2+4)/2c^2\rho^2\sigma^2$, $z=\beta^{1/3}u$
\begin{eqnarray}\label{95}
H(z) = e^{z^3/6}\psi(z)
\end{eqnarray}
with $\psi(z)$ needs to be determined using an iterative procedure, then substitute eq.(\ref{95}) into eq.(\ref{94}), we obtain
\begin{eqnarray}\label{96}
\frac{d^2\psi}{dz^2} +\lambda_0(z)\frac{d\psi}{dz} -s_0(z)\psi = 0
\end{eqnarray}
with
\begin{eqnarray}\label{97}
\lambda_0(z) = \frac{z}{\sigma^2\beta^{2/3}\rho} +z^2
\end{eqnarray}
and
\begin{eqnarray}\label{98}
s_0(z) = \frac{k^2}{\beta^{2/3}\rho^2} -\frac{z^3}{2\sigma^2\beta^{2/3}\rho} -\frac{1}{4}z^4
\end{eqnarray}
Using the MacLaurin series for the first two terms, eq.(\ref{97}) and eq.(\ref{98}) respectively can be reduced to
\begin{eqnarray}\label{99}
\lambda_0(z) = \frac{z}{\sigma^2\beta^{2/3}\rho}
\end{eqnarray}
and
\begin{eqnarray}\label{100}
s_0(z) = \frac{k^2}{\beta^{2/3}\rho^2}
\end{eqnarray}
Substitute eq.(\ref{99}) and eq.(\ref{100}) into eq.(\ref{96}), we obtain
\begin{eqnarray}\label{101}
\frac{d^2\psi}{dz^2} +\frac{z}{\sigma^2\beta^{2/3}\rho}\frac{d\psi}{dz} -\frac{k^2}{\beta^{2/3}\rho^2}\psi = 0
\end{eqnarray}
If we let $s=z/\sqrt{2\sigma^2\beta^{2/3}\rho}$, eq.(\ref{101}) can be written as
\begin{eqnarray}\label{102}
\frac{d^2\psi}{ds^2} +2s\frac{d\psi}{ds} -\frac{2\sigma^2}{\rho}k^2\psi = 0
\end{eqnarray}
Then, using eq.(\ref{14}), it can be obtained
\begin{eqnarray}\label{103}
\lambda_ms_{m-1}-\lambda_{m-1}s_m =\Pi_{n=0}^{m}\left(\frac{2\sigma^2}{\rho}k^2 -2n\right);~~~m=1,2,3,4,...
\end{eqnarray}
According to these conditions, the eigenvalues of energy can be obtained as follows
\begin{eqnarray}\label{104}
\frac{2\sigma^2}{\rho}k^2 = 2n\rightarrow k_n =\frac{\sqrt{\rho n}}{\sigma};~~~n=0,1,2,3,..
\end{eqnarray}
Using eq.(\ref{104}), then eq.(\ref{102}) can be written as
\begin{eqnarray}\label{105}
\frac{d^2\psi}{ds^2} +2s\frac{d\psi}{ds} -2n\psi=0;~~~n=0,1,2,3,..
\end{eqnarray}
Eq.(\ref{105}) is known as the Complementary Error Function Differential Equations. The approximate solution of eq.(\ref{93}) is thus
\begin{eqnarray}\label{106}
H_n(s) = e^{\frac{\omega^2\sigma(\sigma^2+4)}{3c^2\sqrt{2\rho}}s^3}[A~erf~c_n(s) +B~erf~c_n(-s)];~~~n=0,1,2,3,...
\end{eqnarray}
with $erf~c_n(s)$ refers to eq.(\ref{42}).

\subsubsection{7th Model}
Define the character of the optical interface gradation as follows
\begin{eqnarray*}
\varepsilon(x) = \varepsilon_0[csch(\rho x)-\coth(\rho x)]
\end{eqnarray*}
\begin{eqnarray}\label{107}
\mu(x) = \mu_0[csch(\rho x) -\coth(\rho x)]
\end{eqnarray}
If we let $z=\frac{1}{2}\rho x$, and substitute eq.(\ref{107}) into eq.(\ref{3}), we obtain
\begin{eqnarray}\label{108}
\frac{d^2H}{dz^2} -\frac{csch(z)~sech(z)}{\rho}\frac{dH}{dz} +\frac{4}{\rho^2}\left[\frac{\omega^2}{c^2}\tanh^2(z) -k^2\right]H = 0
\end{eqnarray}
In the solution domain is close to zero, the functions of $\tanh(z)$ and $csch(z)sech(z)$ can be approximated by a MacLaurin series in the first three terms, and follow the same way of eq.(\ref{60})-eq.(\ref{62}), then eq.(\ref{108}) can be reduced to
\begin{eqnarray}\label{109}
\frac{d^2H}{dz^2} -\frac{z}{\sigma^2\rho}\frac{dH}{dz} +\frac{4}{\rho^2}\left(\frac{\omega^2}{c^2}z^2 -k^2\right)H = 0
\end{eqnarray}
at $-\sigma\leq z\leq\sigma$ with $\sigma=1$. If we let $s=(4\omega^2/\rho^2c^2)^{1/4}$, eq.(\ref{109}) can be written as
\begin{eqnarray}\label{110}
\frac{d^2H}{ds^2} -\lambda_0(s)\frac{dH}{ds} -s_0(s)H = 0
\end{eqnarray}
with
\begin{eqnarray}\label{111}
\lambda_0(s) = \frac{c}{2\omega\sigma^2}s
\end{eqnarray}
and
\begin{eqnarray}\label{112}
s_0(s) =\frac{2c}{\rho\omega}k^2 -s^2
\end{eqnarray}
Eq.(\ref{111}) and eq.(\ref{112}) can be reduced using the MacLaurin series for the first two trems, so that eq.(\ref{110}) can be written as
\begin{eqnarray}\label{113}
\frac{d^2H}{ds^2} -\frac{c}{2\omega\sigma^2}s\frac{dH}{ds} -\frac{2c}{\rho\omega}k^2H = 0
\end{eqnarray}
Furthermore, if we let $r=s\sqrt{c/4\omega\sigma^2}$, eq.(\ref{113}) can be written as
\begin{eqnarray}\label{114}
\frac{d^2H}{dr^2} -2r\frac{dH}{dr} -\frac{8\sigma^2}{\rho}k^2H = 0
\end{eqnarray}
Next, using the eq.(\ref{14}), we obtain
\begin{eqnarray}\label{115}
\lambda_ms_{m-1} -\lambda_{m-1}s_m =\Pi_{n=0}^{m}\left(\frac{8\sigma^2}{\rho}k^2 +2n\right);~~~m=1,2,3,4,..
\end{eqnarray}
According to these conditions, the eigenvalues of energy can be obtained as follows
\begin{eqnarray}\label{116}
\frac{8\sigma^2}{\rho}k^2 = -2n\rightarrow k_n=\frac{i}{2\sigma}\sqrt{\rho n};~~~n=0,1,2,3,...
\end{eqnarray}
Using eq.(\ref{116}), eq.(\ref{114}) can be written as
\begin{eqnarray}\label{117}
\frac{d^2H}{dr^2} -2r\frac{dH}{dr} +2nH =0;~~~n=0,1,2,3,...
\end{eqnarray}
Eq.(\ref{117}) is known as Hermite differential equation. The approximate solution of eq.(\ref{108}) is
\begin{eqnarray}\label{118}
H_n(r) = (-1)^ne^{r^2}\frac{d^n}{dr^n}\left(e^{-r^2}\right);~~~n=0,1,2,3,...
\end{eqnarray}

\subsubsection{8th Model}
Define the character of the optical interface gradation as follows
\begin{eqnarray*}
\varepsilon(x) = -\varepsilon_0[csch(\rho x)+\coth(\rho x)]
\end{eqnarray*}
\begin{eqnarray}\label{119}
\mu(x) = -\mu_0[csch(\rho x)+\coth(\rho x)]
\end{eqnarray}
If we let $u=\rho x$, and substitute eq.(\ref{119}) into eq.(\ref{3}) we obtain
\begin{eqnarray}\label{120}
\frac{d^2H}{du^2} +\frac{csch(u)}{\rho}\frac{dH}{du} +\frac{1}{\rho^2}\left\{\frac{\omega^2}{c^2}[csch(u) +\coth(u)]^2-k^2\right\}H=0
\end{eqnarray}
In the solution domain is close to zero, the functions of $csch(u)$ and $csch(u) + \coth(u)$ are reduced in the same way as in eq.(\ref{28}) and eq.(\ref{78}), so that eq.(\ref{120}) can be written as
\begin{eqnarray}\label{121}
\frac{d^2H}{du^2} +\frac{u}{\sigma^2\rho}\frac{dH}{du}+\frac{1}{\rho^2}\left[\frac{\omega^2}{c^2}\left(\frac{\sigma^2+4}{2\sigma^2}\right)^2u^2 -k^2\right]H = 0
\end{eqnarray}
at $-\sigma\leq u\leq \sigma$ with $\sigma=1$. If we let $\gamma=\omega^2(\sigma^2+4)^2/\rho^2c^2(2\sigma^2)^2$, and $z=\gamma^{1/4}u$, then eq.(\ref{121}) can be written as
\begin{eqnarray}\label{122}
\frac{d^2H}{dz^2} +\lambda_0(z)\frac{dH}{dz} -s_0(z)H = 0
\end{eqnarray}
with
\begin{eqnarray}\label{123}
\lambda_0(z) = \frac{z}{\sigma^2\gamma^{1/2}\rho}
\end{eqnarray}
and
\begin{eqnarray}\label{124}
s_0(z) = \frac{k^2}{\rho^2\gamma^{1/2}} -z^2
\end{eqnarray}
Eq.(\ref{123}) and eq.(\ref{124}) can be reduced using the MacLaurin series for the first two terms, so that eq.(\ref{122}) can be written as
\begin{eqnarray}\label{125}
\frac{d^2H}{dz^2} +\frac{z}{\sigma^2\gamma^{1/2}\rho}\frac{dH}{dz} -\frac{k^2}{\rho^2\gamma^{1/2}}H = 0
\end{eqnarray}
Next if we let $s = z/\sqrt{2\sigma^2\gamma^{1/2}\rho}$, then eq.(\ref{125}) can be written into
\begin{eqnarray}\label{126}
\frac{d^2H}{ds^2} +2s\frac{dH}{ds} -\frac{2\sigma^2}{\rho}k^2H=0
\end{eqnarray}
Next, using eq.(\ref{14}), it can be obtained
\begin{eqnarray}\label{127}
\lambda_ms_{m-1} -\lambda_{m-1}s_m = \Pi_{n=0}^{m}\left(\frac{2\sigma^2}{\rho}k^2 -2n\right);~~~m=1,2,3,4,...
\end{eqnarray}
According to these conditions, the  eigenvalues of energy can be obtained as follows
\begin{eqnarray}\label{128}
\frac{2\sigma^2}{\rho}k^2 = 2n\rightarrow k_n=\frac{\sqrt{\rho n}}{2\sigma};~~~n=0,1,2,3,..
\end{eqnarray}
Using eq.(\ref{128}), eq.(\ref{126}) can be written as
\begin{eqnarray}\label{129}
\frac{d^2H}{ds^2} +2s\frac{dH}{ds} -2nH = 0;~~~n=0,1,2,3,..
\end{eqnarray}
Eq.(\ref{129}) is known as the Complementary Error Function Differential Equations. The approximate solution of eq.(\ref{120}) is thus 
\begin{eqnarray}\label{116}
H_n(s) = A~erf~c_n(s) +B~erf~c_n(-s);~~~n=0,1,2,3,...
\end{eqnarray} 
with $erf~c_n(s)$ refers to eq.(\ref{42}).

\section{RESULTS AND DISCUSSION}
It has been shown through the eight models that AIM can be used to find approximate solution of electromagnetic wave propagation in the positive-negative gradation of NRM. Some materials that are mathematically can be described as a function which has a singular point has been approached by a linear
function in the domain such that pretty close to the singular point.

\section{CONCLUTION}
This paper has developed an approach for analyzing the distribution of the magnetic field and electromagnetic wave-vector in the positive-negative gradation NRM constructed based on the AIM. As a test of this method, has been used as an example of vacuum-space calculation. Differential equation approach results from the eight NRM models can be solved by Hermite polynomials and the Complementary Error Function.

\section{ACKNOWLEDGMENTS}
ASH grateful to beloved mother, Siti Ruchanah, for her great support. Thanks to Dr Miftachul Hadi for nice discussions and good motivation. Thanks to Professor Muhaimin for very helpful advice. Thanks to Mrs. Yayuk, Dr Imron and Dr Yuni for their kindness. Thanks to Mr. Sjaifuddin Achmad for his inspiring discussion.

\end{document}